# USING THE FRACTAL ATTRITION EQUATION TO CONSTRUCT A METAMODEL OF THE MANA CELLULAR AUTOMATON COMBAT MODEL

**M. K. Lauren, J. M. Smith, J. Moffat and N. D. Perry**

**November 2005**

## Abstract:

In this report the authors examine the possibility of using the Fractal Attrition Equation as a metamodel to describe outcomes of cellular automaton combat models. This Fractal Attrition Equation has been proposed as a replacement for the Lanchester equation. Unlike Lanchester's equation, it is based on a measurement of the spatial distribution of forces involved in a battle, and incorporates this by use of the concept of the fractal dimension. In this report, the form of the expected Loss-Exchange-Ratio function based on this equation is determined. It is shown how this function can be used as a Metamodel to describe the outcome of some cellular automaton models. While this is not an exhaustive proof of the validity of the equation, the case discussed does display interesting features which should be further investigated. This is particularly so given the earlier work of the authors in demonstrating that the Fractal Attrition Equation explains certain features of historical combat data.

EXECUTIVE SUMMARY

**Background**

New Zealand and the UK have been collaborating closely on the development of a new attrition equation for warfare. This equation is different from the standard approach developed by Lanchester because it incorporates spatial patterns into the analysis. This in turn may allow for the better representation of the value of command and control and networked forces in attrition algorithms. It is envisaged that this work may lead to the replacement of many ideas in existing aggregated combat models.

This report has been produced as part of a Project for TP3 of the JSA Group to further investigate this new theory.

**Sponsor**

TP3, JSA Group, TTCP.

**Aim**

To present and demonstrate the validity and usefulness of the Fractal Attrition Equation, and to demonstrate its ability to be used as a Metamodel for the MANA cellular automaton combat model. It is hoped that this work will make a significant contribution to the scientific community and literature in the area of aggregated models of combat.

**Results**

Experiments with simulations suggest that the proposed Fractal Attrition Equation provides a useful metamodel for the behaviour of cellular automaton combat models. In particular, it explains why the power spectrum of the time series of casualties should be a power law.

What was particularly interesting for the scenario discussed in detail here was that the experimental runs revealed that the LER maintained a value of around 1 for several different force ratios. It was also seen that the fractal dimension of each side changed as this force ratio changed, as would be expected according to the Fractal Attrition Equation if the LER was to remain constant.

While the Fractal Attrition Equation was capable of describing this phenomenon, the experimental results were at odds with the classical Lanchester theory. According to this, the LER ratio should change as the force ratio changes, because the potential firepower of one side has changed relative to the other. The Lanchester theory fails to describe the observed behaviour because it does not incorporate spatial distributions.

Furthermore, the Fractal Attrition Equation appears to be consistent with the form of the function describing attacker casualties derived from examination of historical data. Importantly, this new theory provides a theoretical underpinning to describe why dispersed forces, guerrilla and swarming tactics can be effective in certain circumstances, as well as provide insight into the potential effects of new military paradigms such as Network Centric Warfare.

CONTENTS

# 1 INTRODUCTION

## 1.1 Background

This report explores new approaches to equations describing the rate of attrition between two opposing forces. This is motivated by the goal of determining improved attrition algorithms which may better capture the value of enhanced command and control and network enabled capabilities.

The traditional approach to attrition algorithms is to use the Lanchester equations:

$$\begin{aligned} \frac{dR}{dt} &= -k_B B(t), \quad R(0) = R_0 \\ \frac{dB}{dt} &= -k_R R(t), \quad B(0) = B_0 \end{aligned} \tag{1}$$

Note that here we use the "square law" form of the Lanchester equation, which represents direct firing.

However, it has been noted by earlier workers that these equations fail to take account of the spatial distribution of the entities involved in the combat. Thus there is no representation of tactics or structural differences between the two sides which may affect the effectiveness of the deployment of each side's troops.

Here we focus on an alternative approach using the concept of fractal dimension. We do this because the fractal dimension of a spatial distribution contains information about the nature of that distribution. Hence there is the potential to include a spatial element in the algorithm.

This alternative approach uses the Fractal Attrition Equation (FAE) originally proposed in 1999 [4]:

$$\left\langle \left| B(t+\Delta t) - B(t) \right|^2 \right\rangle \propto k_R^{\ 2q} \ \Delta t^{2r}, \quad r = q \tag{2}$$

where $B$ and $R$ are the number of Blue and Red combatants, $k_R$ is the maximum killing potential of a Red combatant, and $q$ and $r$ are unknown functions of $D_R$, the fractal dimension of the distribution of the Red side.

Despite the superficial similarity to the Lanchester equation, the formulation is quite different. Here, the angled brackets represent an ensemble average, so that the equation is statistical rather than a differential equation.

Equation 2 hints at a neat phenomenological model of combat as a self-organising system. It may be imagined that combat fronts (the interface between two groups of combatants) become distorted during the events of the battle. The system of combatants can be viewed as neither rigidly ordered nor completely disordered during this period. The dynamics are in at least some part driven by short-range rules of interaction between individuals rather than via some top-down command process.

Such dynamics are referred to as self-organising, and allow the combat system to be adaptable without completely losing order.

It is expected that, as a result of this self-organisation, the front becomes distorted in such a way that it is most appropriately described by a fractal (noting that most self-organising systems exhibit fractal patterns). Hence the degree of distortion of the front can be expressed as a fractal dimension. In such a scheme, a non-integer value for the fractal dimension indicates that the distribution is fractal, and its value indicates the degree to which the front is an "irregular" curve.

The equation is therefore intended to describe the rate of attrition during a period in which the $B$ Blue and $R$ Red combatants are constantly within close enough range of each other to self-organise as a result of the other side's presence.

While this makes a nice phenomenological model, the aim of this work is to explore methods for obtaining more quantitative predictions of combat behaviour from Equation 2. This is done by comparing the expected theoretical Loss Exchange Ratio derived from Equation 2 with the outcomes of a simple cellular automaton combat model.

Previous investigations by the authors [8, 11-16] have demonstrated that some of the features one might expect to find in combat data if Equation 2 holds do indeed exist. In particular, it has been demonstrated [8,15] that the power spectrum of casualties from historical battles exhibit the power-law nature that the equation implies. This is a very strong prediction from the metamodel, and relates directly to the volatility of casualty time series (that is, they should be intermittent in nature, rather than smoothly varying).

Furthermore, it has been shown with cellular automata models, particularly the MANA and ISAAC models, that combatants tend to self-organise into fractal-type patterns, and that under these circumstances the rate of attrition can appear to behave as a power-law function of $k$ [15]. Moffat and Witty have further shown that fractal analysis appropriately describes the distributions of troops during exercises.

However, this previous work, while supporting the power-law nature of Equation 2, has so far failed to demonstrate conclusively the relationship between attrition rate and the power exponent $q(D_R)$, or even between $q$ and $D_R$.

Here it will be shown how the expected Loss-Exchange Ratio (LER, defined as the number of Blue casualties over the number of Red casualties) appears to be dependent on $D_R$, at least for the cellular automaton model examined here. A formulation is produced which estimates the LER in terms of the fractal dimension of the distribution of the sides. This formulation thus acts as a Metamodel for the more complex cellular automaton model used.

### 1.2 Spectra of casualty data

Before examining how the LER function can be obtained in terms of the fractal dimension, $D$, it is worth discussing approaches one might consider to measuring the parameters in Equation 2 directly.

As noted in the introduction, previous investigations into the relationships between $q$, $r$, $D$ and attrition rate have not been fruitful. Rather, it has simply been shown that analysis of historical data and model simulations are not inconsistent with Equation 2.

Here we briefly discuss these difficulties. In principle, it is possible to measure the values for $q$ and $r$ directly. However, measurement of parameters for Equation 2 is not as straightforward as it might be for a differential equation such as Lanchester's equation, due to its statistical nature.

One approach is to recognise Equation 2 as a second-order statistical moment of a casualty data time series, and attempt to estimate the new value $r$, and hence $q$. To do this we can use the Wiener-Khinchine relation which states that the Power Spectrum of a time series is the Fourier Transform of the autocorrelation of the series. As we will see later, this implies a Power Spectrum of the form;

$$P(f) \propto f^{-\beta} \qquad (3)$$

Equation 3 has been observed to hold for both historical and simulated casualty data [8, 15], and examples will be shown in detail later in the Report. One might then expect that it should be a simple matter to measure the value of $\beta$ and hence obtain $r$ and $q$, then to compare these values with both the attrition rate and the fractal dimension of the distribution of each side. The prediction of our metamodel that casualties have a power law power spectrum is a strong prediction, and directly relates to the nature of the time series themselves. It predict that they will be intermittent ('bursty') in nature, rather than smoothly varying. We will also discuss later how this links to estimation of the volatility of the time series of casualties.

In practice it proves difficult to robustly estimate $\beta$. There are two reasons. The first is that it is not straightforward to estimate $\beta$ from a plot of the spectrum. Recalling that on a log-log plot a spectrum of the form of Equation 4 will display a straight line, the slope of which is $\beta$, it is typically found for casualty data that the region of the spectrum which obeys this law can be relatively small, and may tend to display a slightly curved nature at the start and finish so that it is ambiguous as to where to measure the slope.

The second problem is that often the time series of casualties is sparsely "populated", particularly for simulations, due to the small number of combatants. For example, in the simulation used later in this report there may be a combined total of 100 casualties in a 200-time step simulation. In this case, the spectra typically either display poorly defined power-laws, or may simply resemble white noise.

While historical data have been demonstrated to display convincing power-law spectra, for that data there were many (often hundreds) of casualties per time step (where a time step is a day). The difficulty with reproducing this within a combat model is in modelling large numbers (thousands) of soldiers fighting complex battles.

A more promising alternative for generating model data is to use "contact" data, as was done in [8]. In this case, the time series describes the number of enemy entities seen per time step rather than casualties. This produces a richer time series, and it may be argued (as indeed we do later in the derivation of our metamodel) that the number of contacts ought to be proportional to number of casualties. However, though this method produces much more convincing power spectra, the connection between the spectral slope $\beta$ and the rate of attrition is still not directly apparent from the experimental data.

Therefore other methods are needed.

### 1.3 Dependence of attrition on *k*

An alternative to trying to determine the value of *r* in Equation 2 by using the power-law exponent $\beta$ from the spectra of casualty time series is to examine how the attrition rate depends on the value of *k*.

As with the spectral analysis approach, this methodology has been tried in previous reports [4, 8, 15]. It involves finding the average time for one side to suffer a certain number of casualties for a given value of *k* for the other side. This average time is then plotted against *k* on a log-log graph to theoretically give a straight line, the slope of which is *q*.

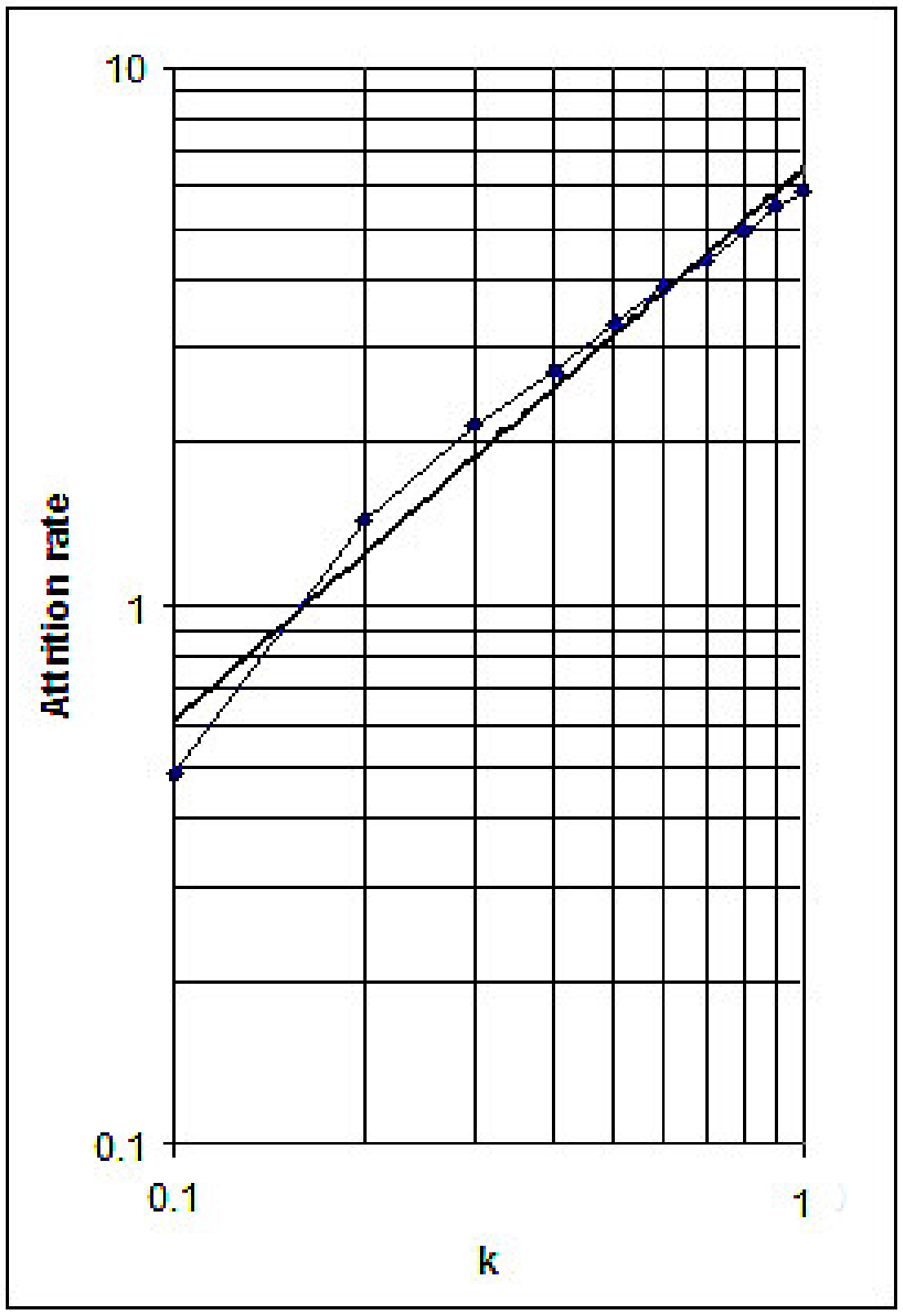

**Figure 1: Rate of attrition as a function of the opposing side's *k*. While the plot approximates the theoretically expected straight line, determining what the slope of this line should be is problematic.**

While this previous research has suggested that the attrition rate does approximate a power-law dependence on $k$, once again there are problems determining $q$ from the slope displayed on these plots. An example is shown in Figure 1.

The principal difficulty is that, as with finding the power-law exponent for the spectrum, it is difficult to decide whereabouts the straight line should be fitted due to slight curvature in the data points.

Note, however, that the existence of this curvature does not necessarily show that a power-law type relationship is not valid, nor does it disprove Equation 2. As has been pointed out previously [7], one would not expect the plot to exhibit a perfectly straight line.

Above some value of $k$, targets are eliminated by shooters almost immediately after contact. In this case, neither force has the opportunity to manoeuvre and self-organise. Hence one might expect the fractal dimension $D$ to have a slightly different value depending on $k$. Additionally, increasing $k$ further has little effect on the attrition rate.

Given both of these points, one would expect that the plot of attrition rate against $k$ ought to curve slightly on a log-log plot, since $D$ is not constant with $k$.

In summary, our attempts to measure $q$ and $r$ directly using these methodologies have not produced useful quantitative results to date.

### 1.4 Determining the fractal dimension *D*

For this report, the fractal dimension of the distribution of combatants was determined by using a box-counting technique. This involves splitting the MANA-model battlefields into four equal squares and counting the number of squares containing either Red or Blue automata, depending on the side for which the fractal dimension is being calculated. Then each of these squares is split into four, with automata-containing squares counted again, and so on. This is illustrated schematically in [6].

The "box-counting" fractal dimension, $D$, is then given by:

$$D = \lim_{d \to 0} \frac{\log N}{\log\left(\frac{1}{d}\right)}$$

where $d$ is the width of the box, and $N$ the number of boxes required to cover all the automata. Note that this is equivalent to finding the slope of a straight line fitted to a plot of $N$ versus $d$ on a log-log plot.

Figure 2 shows an example of such a plot using data generated from the MANA model, using a spatial distribution of automata similar to that shown for the Meet scenario in the later sections. As can be seen from the figure, in this case the slope was determined by fitting a straight line to points 2 to 5. As with the discussions in the preceding sections on fitting straight lines to spectra and for attempting to directly determine the value of $q$, it is not obvious from Figure 2 which points to include in the fit. This is particularly so since there appears to be a “knee” in the plot at about point 5, so that for the points beyond this, the slope becomes substantially less steep.

Generally one would expect a “truncated” range of scales for which this power-law behaviour applies. At the smallest scales, the size of the boxes have become smaller than the characteristic distance between the automata. Hence all the automata already lie individually in boxes, so that reducing the box size further does not increase *N*. For the largest scales, a single box covers all automata, hence increasing box size further also has no effect on *N*.

However, while this difficulty in fitting a line to a log-log plot was problematic in the context of determining *q* and *r*, it was found that finding the “exact” fractal dimension was not so critical for estimating the expected LER. Rather, what was important was to be consistent about the method for estimating *D*. Thus if the estimate for $D_R$ was based on a fit of the 2nd through to the 6th points, then so too should the estimate for $D_B$.

Note that the value of *D* used in the Metamodel estimations of LER was in fact *D* averaged over all time steps. That is to say, *D* was calculated at each time step and the average value used.

In the following section, it will be shown how these estimates for *D* were able to be used in conjunction with Equation 2 to construct a Metamodel describing expected LERs for at least one MANA model scenario.

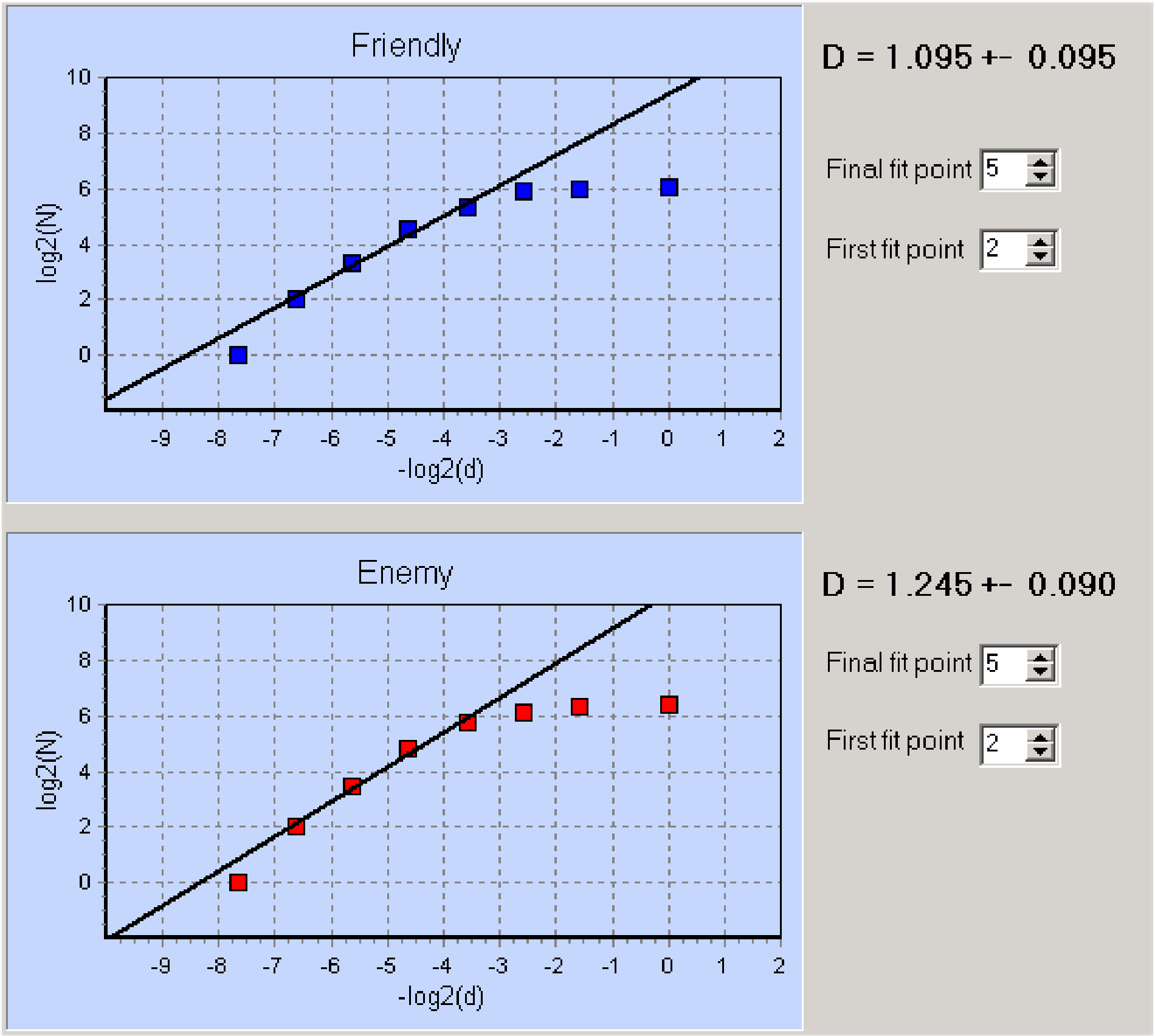

**Figure 2: Determining the fractal dimension by box-counting techniques involved finding the slope found by plotting the number of boxes containing automata (*N*) against the size of the box (*d*).**

## 2 FRACTAL ATTRITION EQUATION

### 2.1 Derivation

In this section we describe the derivation of Equation 2 and show how this may be used to formulate an expression for the Loss-Exchange-Ratio between two groups of combatants. In later sections, we will explore the applicability of this formulation to the outcomes of the MANA cellular automaton model.

The first part of our analysis is heuristic in nature and is used to motivate the selection of the key factors which constitute the Fractal Attrition Equation (FAE).

Consider a Blue agent in a cellular automata model of conflict, travelling at a velocity $v$ for a time $\Delta t$. Let $d = v\Delta t$ be the distance travelled.

If the Red force is fractally clustered, then by definition of the fractal dimension $D$ of these forces, the probability of encountering a Red cluster in this box of width $d$ is $\propto (v\Delta t)^{-D}$ since this is directly related to the probability of this box being occupied by one or more Red agents. If now $n_d$ = mean number of Red agents in an occupied box of side $d$, then the number of encounters with Red agents is $\propto (v\Delta t)^{-D} n_d$. We assume that we can approximate $n_d$ by the total Red force size $R(t)$ divided by the number of boxes. Neglecting constants (the length and width of the battlefield), this gives the expression $n_d \propto R(t)(\Delta t)^2$. We then have that the number of encounters by a Blue agent is given by an expression of the form $(\Delta t)^{2-D} R(t)$.

The expected attrition rate of Blue agents is then given by:

$$E_{0<t<T}\left(\left|\frac{\Delta B}{\Delta t}\right|\right) \propto \text{(unit kill probability of Red) x (Probability of meeting a cluster in time period } \Delta t \text{) x Red force size}$$

where $E$ denotes the expectation over the time period $(0,T)$.

Thus we assume the functional relationship:

$$E_{0<t<T}\left(\left|\frac{\Delta B}{\Delta t}\right|\right) = E_{0<t<T}\,\varphi(k, (\Delta t)^{2-D}, R(t)) \qquad (4)$$

in terms of the key driving variables of the system. Here $B$ is the Blue force size at time $t$, $k$ is the agent unit effectiveness, $\Delta t$ is the timestep of interest and $R(t)$ is the Red force size at time $t$. Note that the expectation in Equation (4) is taken over a number of such timesteps $\Delta t$ spanning the total time period $T$ (we have dropped the variable $v$ since this is assumed constant).

Applying the approach of scale invariant metamodelling ([11] Chapter 4), we then set up the Fundamental Gauge Class: F = Force, T = Time and L = Length. In terms of these dimensions, we have:

$$
\begin{aligned}
&[\frac{\Delta B}{\Delta t}] = FT^{-1} \\
&[k] = FT^{-1}F^{-1} = T^{-1} \\
&[(\Delta t)^{2-D}] = T^{2-D} \\
&[R(t)] = F
\end{aligned}
$$

Equating[1] dimensions on both sides of the relationship, we can write the Fractal Attrition relationship in dimensional terms as [6]:

$$
\begin{aligned}
FT^{-1} &= [k]^{\alpha}[\Delta t^{2-D}]^{\beta}[R(t)]^{\gamma} \\
&= T^{-\alpha}T^{\beta(2-D)}F^{\gamma}
\end{aligned}
$$

Hence

$$
\begin{aligned}
&\gamma = 1 \\
&\beta(2-D) = \alpha - 1
\end{aligned}
$$

and we have the relationship:

$$
E_{0<t<T}\left(\frac{\Delta B}{\Delta t}\right) \propto k^{\alpha}\Delta t^{\alpha-1}R(t) \qquad (5)
$$

However, the asymptotic approximation technique used to obtain this equation is strictly only correct when all parameters dimensionally balance. In the present case, $k$ and $\Delta t$ both depend on time. However, $\Delta t$ was introduced as a substitute for $d = v\Delta t$, on which the heuristic explanation properly depends. The parameter $d$ can be kept until after the asymptotic approximation technique has been applied and then the substitution for $d$ can be made, resulting in equation 5 as before.

Experimental evidence [6], [18] suggest that:

$$
\alpha = \frac{D}{2}
$$

Thus:

$$
\beta(2-D) = \frac{D}{2} - 1 = -\frac{1}{2}(2-D).
$$

[1] The heuristic approach described here is consistent with the formal development of a metamodel of Type 2 (as in Reference 11 Chapter 4) where $E(\frac{\Delta B}{\Delta t}) \propto kR\Phi(k\Delta t)$ and the renormalisation invariant function $\Phi$ has the form $\Phi(x) = x^{\xi}$ with $\xi$ an 'anomalous dimension', as usual where fractal systems are involved. (See G I Barenblatt 'Scaling, Self-Similarity and Intermediate Asymptotics', Cambridge University Press, Cambridge, UK, 1996, Chapter 12). In our case, $\xi = -(1-\frac{D}{2})$.

From which we derive $\beta = -\frac{1}{2}$ and $\alpha - 1 = \beta(2-D) = -(1-\frac{D}{2})$.

When there is no dispersed fractal clustering, and Red agents form an uncorrelated set spread over the battlespace, with Fractal Dimension $D$ = 2, we require that Equation 5 reverts to the square law Lanchester equation, i.e.:

$$E_{0<t<T}\left(\frac{\Delta B}{\Delta t}\right) \propto kR(t)$$

This occurs when $\alpha = 1$, which also eliminates the term in $\Delta t$ as required.

### 2.2 General form and Loss Exchange Ratio

We thus assume that the FAE relationship takes the general form:

$$E_{0<t<T}\left(\frac{\Delta B}{\Delta t}\right) = E_{0<t<T}\left(-ck^{\frac{D}{2}}\Delta t^{-(1-\frac{D}{2})}R(t)\right)$$

where $c$ is the constant of proportionality.

Note that $D$ and $R$ are both functions of time $t$ here.

The Loss Exchange Ratio (LER) for the two forces Red *(R)* and Blue *(B)* is then given by the following:

$$\begin{aligned} LER &= \frac{E_{0<t<T}(\Delta R)}{E_{0<t<T}(\Delta B)} \\ &= \frac{E_{0<t<T}(\Delta R/\Delta t)}{E_{0<t<T}(\Delta B/\Delta t)} \\ &= \frac{E_{0<t<T}\left(-ck_B^{D_B/2}\Delta t^{-(1-D_B/2)}B(t)\right)}{E_{0<t<T}\left(-ck_R^{D_R/2}\Delta t^{-(1-D_R/2)}R(t)\right)} \\ &= \frac{E_{0<t<T}\left((k_B\Delta t)^{D_B/2}B(t)\right)}{E_{0<t<T}\left((k_R\Delta t)^{D_R/2}R(t)\right)} \end{aligned} \qquad (6)$$

### 2.3 Correlation in Time

Though we will not use the following in the analysis of the simulated combat discussed in the later sections, it is interesting to note the expected statistical behaviour of combat data if Equation 5 holds, and how this can be related back to Equation (2).

We assume that the combat system is similar to a turbulent flow system where turbulence is multiscaling (in the sense that the effect is recursive at different levels of system resolution). This leads to the assumption of what are called *multifractal statistics.* This means that the form of expression above applies in an analogous way to all of the moments of the random variable $\left|\frac{B(t+\Delta t)-B(t)}{\Delta t}\right|$. Thus we have:

$$E_{0<t<T}(\left|\frac{B(t+\Delta t)-B(t)}{\Delta t}\right|^{p}) \propto \Delta t^{g(D,p)}$$

where $g$ is a function only of the Red force fractal dimension $D$ and the order of the moment $p \geq 1$.

In particular, (noting that $\Delta t$ is a constant relative to the expectation $E$ over time $t$), the second moment ($p$=2) should have the relationship;

$$E_{0<t<T}(\left|\frac{B(t+\Delta t)-B(t)}{\Delta t}\right|^{2}) = \frac{1}{\Delta t^2} E(\left|B(t+\Delta t)-B(t)\right|^2) \propto \Delta t^{g(D,2)}$$

and hence:

$$E_{0<t<T}(|B(t+\Delta t)-B(t)|^2) \propto \Delta t^{g(D,2)+2} \qquad (7)$$

which is the essential assumption of Equation (2).

We recall that in this equation, *B(t)* is the number of Blue casualties at time *t*, *E* denotes expectation over the time period *T,* and *D* is the fractal dimension of the Red force locations on the battlefield, corresponding to local clustering of Red agents. In this way, clustering in time of Blue casualties is related to clustering in space of Red forces on the battlefield.

If we expand the square term on the left hand side of Equation 7, we have:

$$\begin{aligned} E_{0<t<T}(|B(t+\Delta t)-B(t)|^2) &= E\left[B(t+\Delta t)^2 - 2B(t)B(t+\Delta t) + B(t)^2\right] \\ &= 2s^2 - 2\text{corr}(B(\Delta t)) \end{aligned}$$

where $s^2$ is the second moment of the random variable *B(t),* and $\text{corr}(B(\Delta t))$ is the auto-correlation for the time-series of Blue casualties *B(t)* with lag $\Delta t$. We assume here that $\Delta t$ is small relative to the complete timespan *T* over which we are taking expectations *E,* with $0 < t < T$.

Assuming also that $s^2$ is constant, we can thus see from Equation 7 that:

$$\text{corr}(B(\Delta t)) \propto \Delta t^{g(D,2)+2}$$

Considering now a range of values of the lag $\Delta t$. It follows from Jensen ([19] page 9 and Appendix D) that:

$$F\{\text{corr}(B(\Delta t))\}(f) \propto |f|^{-(g(D,2)+2+1)} = |f|^{-(g(D,2)+3)} \qquad (8)$$

where *F* is the Fourier Transform.

Let B*B denote (in functional terms), the auto-correlation of the time series *B*, where we think of *B* as a function of time *t*. From standard Fourier Transform theory, we have that:

$$F(B*B)(f) = F(B)(f)\bar{F}(B)(f) = |F(B)(f)|^2$$

where again *F* is the Fourier Transform function, and *f* a specific frequency value. Thus the Fourier transform *F* of the auto-correlation function is just the square of the Fourier Transform of the time series *B(t)*. This is, by definition, the Power Spectrum of the Blue casualty time series *B(t)*.

It thus follows that:

$$|F(B)(f)|^2 = F\{corr(B(\Delta t))\}(f) \propto |f|^{-(g(D,2)+3)} \qquad (9)$$

Putting this together, the left hand side of Equation 9 is the Power Spectrum of the time series of Blue casualties *B(t)*. According to Equation 9, this should vary with frequency *f* in accordance with a power law, where the exponent of the power law is a function of the fractal dimension *D* of the Red force, corresponding to the agile clustering of the Red agents.

The simplest assumption we can make is that *g(D,2)+2=D*, which leads to the relation

$$\left\langle |B(t+\Delta t) - B(t)|^2 \right\rangle \propto \Delta t^D$$

from Equation 7, and was initially suggested in [10].

This further implies that the Power Spectrum of the time series of Blue casualties has the form

$$|F(B)(f)|^2 \propto |f|^{-(D+1)}.$$

However the precise form of the power law exponent has to be determined experimentally.

# 3 COMPARISON WITH MODEL DATA

## 3.1 The MANA cellular automaton model

In this section we explore the applicability of the theory derived in Section 2 to describing the outcomes of abstract combat models constructed within the MANA (Map Aware Non-uniform Automata) cellular automaton model (Version 3), developed by the Defence Technology Agency, New Zealand [9].

Within this model, automata choose their moves according to personality weightings in a similar way to the ISAAC model described by Ilachinski [3]. The movement algorithm works by calculating the penalties associated with an automaton moving to any of the cells surrounding it [1]. The move is chosen by randomly selecting a cell with a penalty lower than some threshold, hence automata do not necessarily make "perfect" moves each turn. Penalties are smallest for moves that bring the automata nearest to objects for which they have the greatest weighting.

The order in which automata move is randomly selected each turn, and no more than one automaton can occupy a given cell. For shooting, no systematic target selection is modelled. Instead, targets are chosen at random from those available. Because automata move one at a time in a defined order, there are no "double kills" where one automaton may end up shooting at an enemy automaton that has already been killed.

For the model scenario discussed here, kills are determined by a simple system where, if a target is within range of an automaton's weapon, then it will be killed with a certain probability.

While the MANA model also allows somewhat sophisticated behaviour to occur by the use of "triggers" and more sophisticated layers of rules, the simple scenario explored in this report does not make use of these. Additionally, parameter settings for the scenario will only briefly be discussed. This is because from the point of view of the Metamodel, it is only important to characterise the positions of the automata as the models evolve, rather than concern ourselves with how the automata got to those positions (i.e. the rules of movement).

## 3.2 Test Scenario

Here a single scenario will be discussed in detail. However, a number of different scenarios were also examined for their consistency with the Metamodel. For the scenarios used, the entities within the model were given relatively low kill probabilities in order to allow enough time for self-organisation to occur before one or other force was eliminated.

The value used was 0.01, meaning in a given time step in which a target automaton is within the shooting range of another automaton, there is a 0.01 chance of the target being killed in that time step.

Other model parameters were set so as to be consistent with the assumptions of the metamodel. The firing and sensing ranges were set to 20 cells, compared to the size of the battlefield of 200x200 cells.

Note that the kill probability used in the MANA model is not the same as $k$ in Equation 1. Generally speaking, there are other parameters which affect the killing rate, such as firing range. For example, in the case where the two forces have uneven firing range, the force with the longer range will have a higher effective kill rate than its opponent even if both have the same killing probability. However, for the scenarios studied in this report, the firing range of both sides was always the same, and $k$ was simply taken to be the single shot kill probability.

This report discusses the results from the "Meet" scenario previously introduced by Witty [17]. Generally speaking, there are a large number of variations possible in scenario set-up, making it impossible to explore all possibilities in this report.

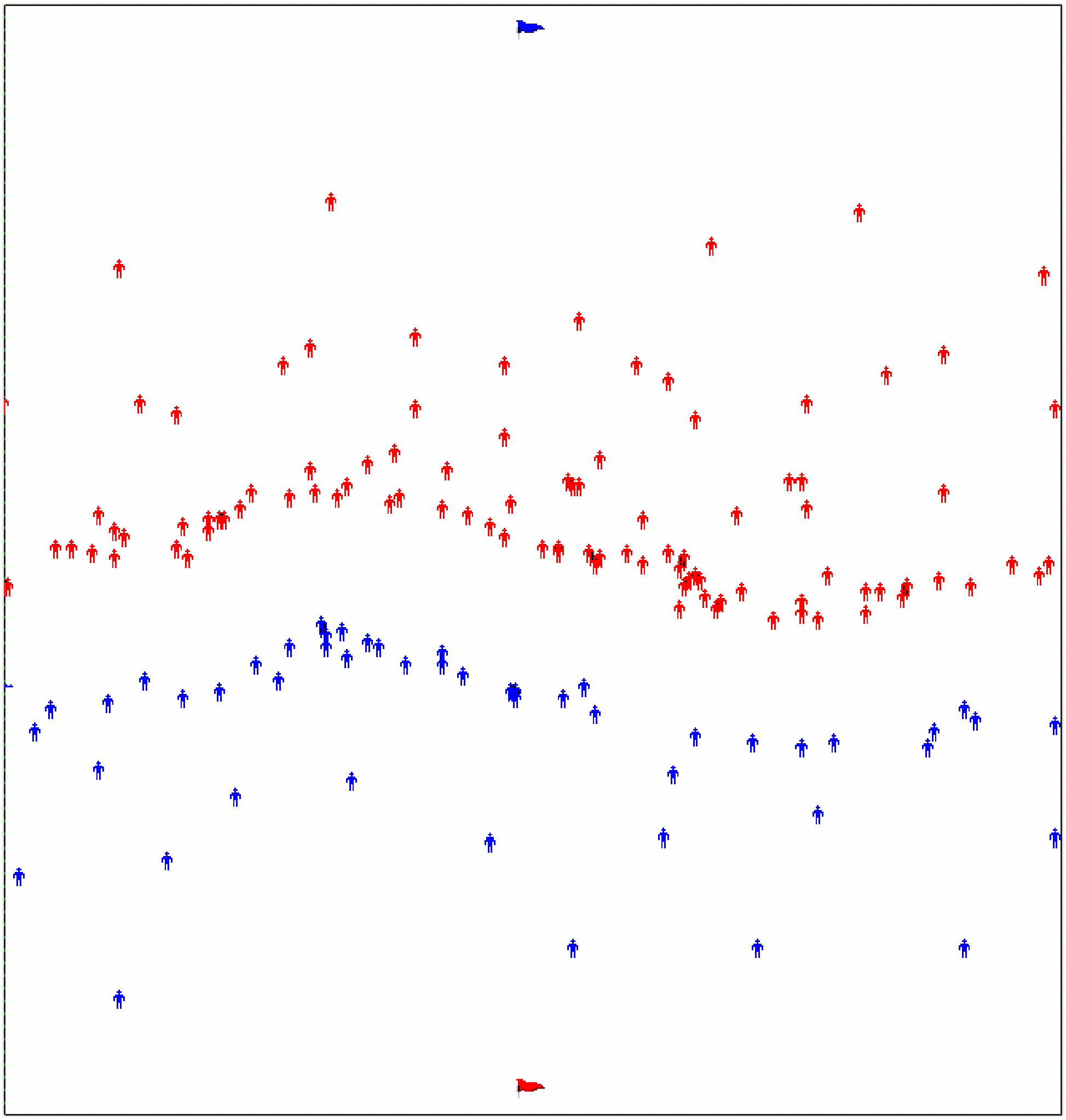

**Figure 3: Screen shot from the Meet scenario, showing the first phase of the run, where both sides form a "line", which displays a wave-like motion.**

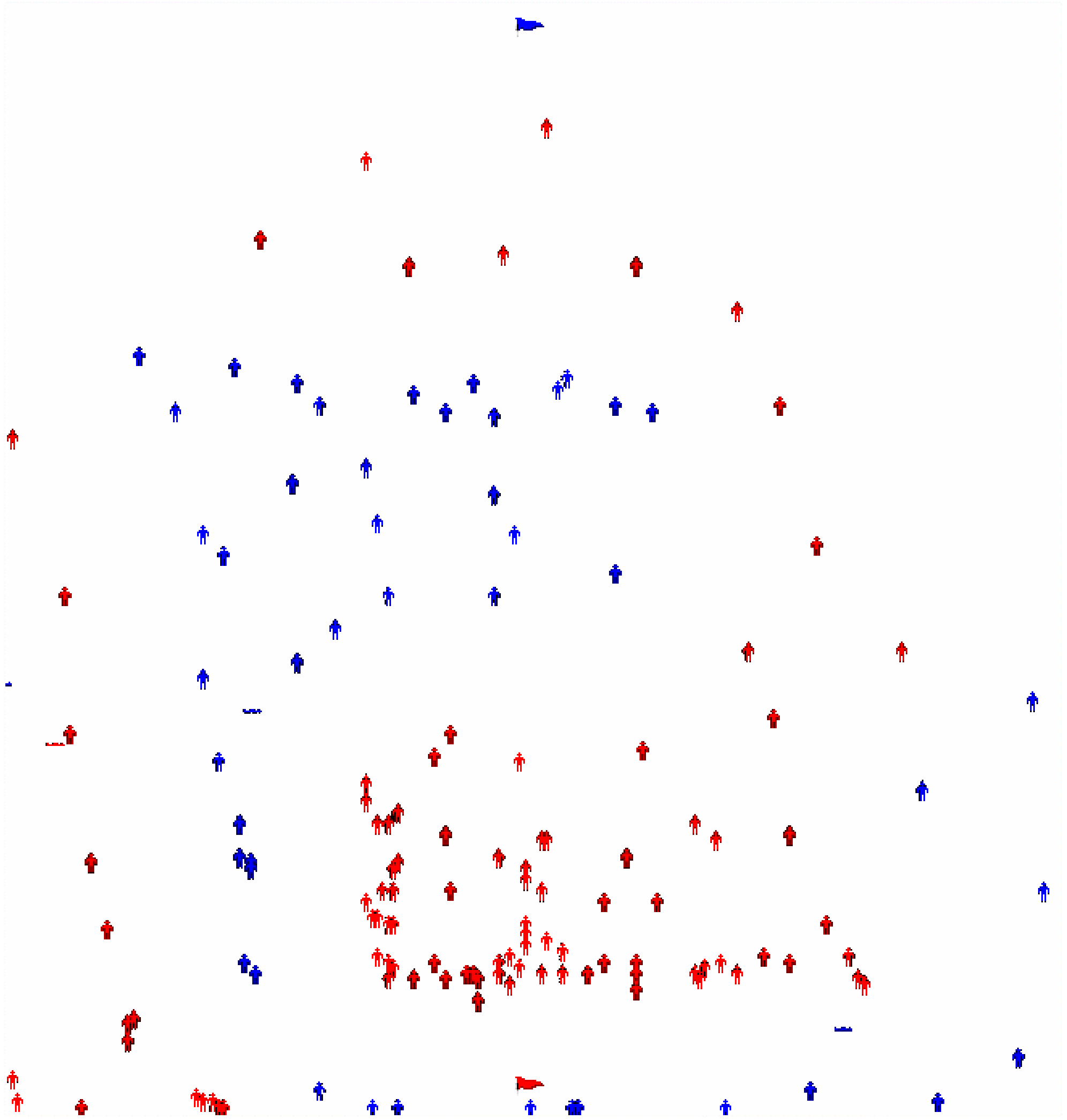

**Figure 4: Screen shot from the Meet scenario, showing the later phases of the same run shown in Figure 3.**

Additionally, how well a scenario is described by the Metamodel will depend on how well it matches its assumptions. For example, for scenarios where automata from one side fail to make a contribution to attrition (if, say, they remain in an area where contact with the enemy never occurs) but nonetheless count towards the fractal dimension estimation, it may be expected that that scenario would be poorly represented by the Fractal Attrition Equation.

The Meet scenario is constructed so that the two sides meet in the centre of the battlespace and initially form a line, which moves in a wave-like manner. However, this line rapidly becomes unstable and breaks up into multiple clusters of automata after some unpredictable period of time. Figures 3 and 4 show the appearance of the distributions of Red and Blue automata at different stages of a single model run of the scenario.

The personalities of the automata are simple. Each automaton is repelled by other automata, and most strongly repelled by enemy automata (a weighting of –80 for enemy and –20 for friends). They are also attracted to move toward their waypoint (flag) at the opposite end of the screen (with a weighting of 10).

Despite the simplicity of these rules, it is apparent from the figures that they are capable of generating complex and generally unpredictable behaviour.

Examination of many different scenarios suggested a relationship between the Loss Exchange Ratio and the fractal dimension that was in fact reasonably well described by an equation of the form:

$$\mathrm{LER} = \frac{R\left(k_R\right)^{D_R/2+1}}{B\left(k_B\right)^{D_B/2+1}} \qquad (10)$$

somewhat different from that expected from the theory (compared this with Equation 6). Here we do not explore reasons for this disparity. Nonetheless, it is important to realise that Equation 10 has a form which is highly consistent with the thrust of the theory presented, and we feel does more to support the theory than undermine it.

Table 1 compares the predictions for LER from both the Lanchester equation from Equation 10 with the actual average outcome of the MANA model, using the Meet scenario.

| | | | | | | | | LER | | | | |
|---|---|---|---|---|---|---|---|---|---|---|---|---|
| | $k_B$ | $k_R$ | $D_B$ | $D_R$ | $B_0$ | $R_0$ | $t$ | **MANA** | Lanchester | Fractal Attrition Equation | Correction factor for Lanchester | Correction factor for FAE |
| Case 1 | 0.01 | 0.01 | 0.82 | 1.05 | 50 | 100 | 200 | **1.1** | 2.3 | 1.2 | 2.1 | 1.1 |
| Case 2 | 0.01 | 0.01 | 0.81 | 1.23 | 30 | 100 | 200 | **1.0** | 5.3 | 1.3 | 5.3 | 1.2 |
| Case 3 | 0.01 | 0.01 | 1.14 | 1.23 | 80 | 100 | 200 | **1.0** | 1.3 | 1.0 | 1.3 | 1.1 |
| Case 4 | 0.03 | 0.01 | 1.22 | 1.38 | 50 | 100 | 150 | **0.42** | 0.68 | 0.24 | 1.6 | 1.8 |

**Table 1** Results from the MANA Meet scenario compared with the loss-exchange ratio expected using the Lanchester equation and the Fractal Attrition Equation.

For the above cases, the Fractal Attrition Equation performed at least as well as the Lanchester estimates. However, for this example the size of the error was not the interesting point. Rather, the results shown here demonstrate how a numerically inferior force (Blue) can be expected to improve its LER by adopting a more favourable distribution (and fractal dimension) than its opponent.

This can be seen since the LER is close to one for three of the cases of the MANA experimental runs, despite the numerical disparity between the two forces. This is not expected according to the Lanchester equation, but is described by the Fractal Attrition Equation in terms of the fractal dimension for each force changing in such a way that Red causes a lower rate of attrition to Blue than Blue causes to Red. Significantly, this is exactly what we see in the behaviour of the distribution of forces, even if the relative size of the error is not much better than the Lanchester calculation in this case.

Examination of the results of other MANA scenarios not discussed in this report suggested that, as a general rule, the Lanchester model tended to work better than the metamodel when the automata maintained a single clump-like formation, or when there was a disparity in $k$ between the two sides.

## 4. COMPARISON WITH HISTORICAL DATA

In this section we discuss the connection between the Fractal Attrition Equation and historically observed outcomes for land battles. Analysis of historical battles suggest that the number of attacker casualties, $C$, is described by [16]:

$$C = \left(\text{Number of attackers/Number of defenders}\right)^{0.685} \qquad (11)$$

where $C$ is a multiplier to the expected number of attacker casualties. That is to say, attacker casualties actually increase as number of attackers increases, the opposite outcome expected from the Lanchester equation. Just as surprising is that the power exponent in this case, 0.685, is a non-integer with a value that cannot obviously be explained (although in [11] it was noted that the fractal dimension of the interface, at 1.685, lies in the correct range as predicted by the cellular automaton process of invasion percolation).

Figure 5 shows evidence that the type of relationship suggested by Equation 11 holds for MANA combat simulations. The data plotted represents the number of attacker casualties for various-sized attacker forces against 60 defenders. The simulation was set up so that the attacking force simply advances in a multiple-rank line-like formation on a static linear formation of defenders, also in multiple ranks.

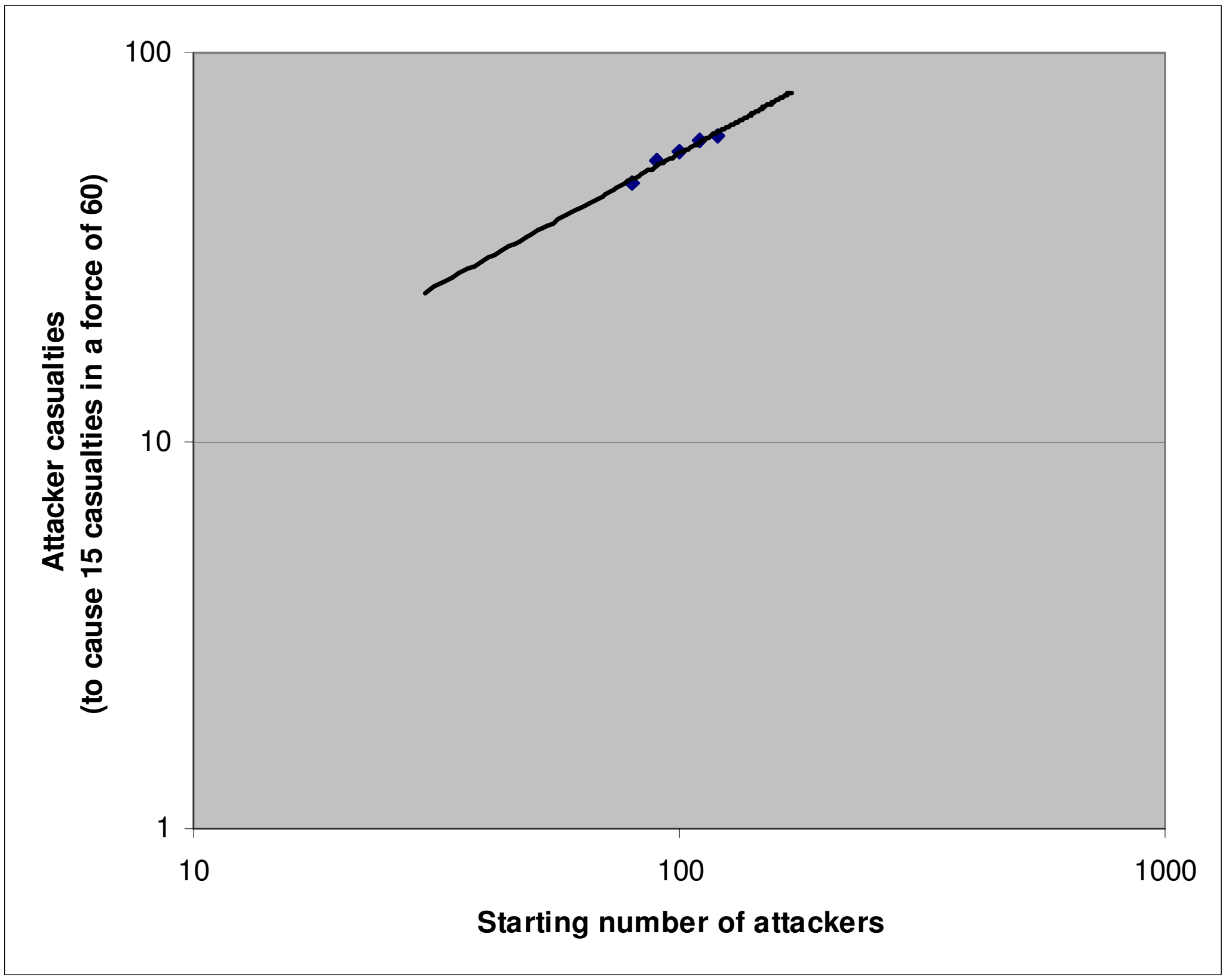

**Figure 5: Number of attacker casualties as a function of starting number of attackers, where the defending force is 60 strong. For the points plotted, the graph is reasonably consistent with a function of the form of Equations 2 and 11. While a very small number of points have been used on this plot, for practical purposes we are only interested force ratios (i.e. the number of attackers divided by defenders) from 1 to about 4, hence the range is appropriate.**

The attackers began beyond the shooting range of the defenders so that, at least in the early time steps, only part of the attacking force is in shooting range.

The simulation was set so that all entities could simultaneously shoot at 10 opponents given the opportunity, and kill them with a probability of 0.5 per target. This was because earlier experiments [10] have shown that individuals need the ability to hit multiple enemies in a volley for $Q$ to be positive. As discussed in Moffat [11], this allows the defender rate of fire to be proportional to the number of attacker targets. With this level of lethality, simulation runs typically lasted 7 or so time steps.

By setting $a = 10$, $k = 5$ and $\Delta t = 7$, one obtains a value for $Q$ of 0.77 from Eq. (15). In this case, $k$ once again represents the maximum killing potential per entity per time step, so that a value of 5 represents that a single entity can kill up to 10 enemy entities, each with a probability of 0.5 per time step. While this value for $Q$ is not very similar to the power exponent value in Eq. (2), the important point to note is that it is a non-negative non-integer.

It can be seen from Figure 5 that, for the values presented, a function of the form of Equation 14 is appropriate, where in this case $Q = 0.68$. Note that while a very small number of points have been used on the plot, for practical purposes we are only interested in force ratios (i.e. the number of attackers divided by defenders) from 1 to about 4. Thus we need only examine a small number of points.

Adding points representing greater or lower force ratios for these simulations undoubtedly destroys the straight-line appearance of the plot. However, Equation 14 can hardly be expected to hold for real situations in the limiting cases of the number of attackers either going to zero or infinity.

## 5 FRACTAL EQUATION OF STATE

We now return to Equation 10. Using the properties of fractals, the relationship between the number of discrete discernable entities in the Red force, *R*', and *D* is described by:

$$R' \propto a^{-D} \quad \Rightarrow \quad D = c - \log_a R' \qquad (12)$$

where $a$ is the resolution at which the battlefield is examined. If we take $a$ to be the size of the smallest area which at most contains a single Red entity, then $R' = R_0$. If we assume that $a$ and $k$ are about the same for the Red and Blue forces, substituting Equation 12 into the right hand side of Equation 10 produces:

$$\frac{R(t)}{B(t)}\left(k\,\Delta\,t\right)^{\log_a (B_0/R_0)/2} \qquad (13)$$

which can be rearranged to give:

$$\frac{R(t)}{B(t)}\left(\frac{B_0}{R_0}\right)^{Q} \qquad (14)$$

where $Q$ has the form:

$$Q = 0.5\log_a\left(k\,\Delta t\right) \qquad (15)$$

Replacing the right hand side of Equation 10 with Equation 14, and cross-multiplying yields:

$$B(t)\Delta B = \left(\frac{B_0}{R_0}\right)^{Q} R(t)\Delta R \qquad (16)$$

Dividing the interval between the beginning and ending of the engagement into a series of such time intervals allows Equation 16 to be applied to each of these time steps. A connection between the initial and final force strengths can then be obtained by summing all those expressions:

$$\sum_{\text{int}\,ervals} B\Delta B = \left(\frac{B_0}{R_0}\right)^{Q} \sum_{\text{int}\,ervals} R\Delta R \qquad (17)$$

In keeping with the use of the FAE as a metamodel for MANA, the force strengths can be seen to represent the number of live entities. When attrition occurs, it reduces that number by integer values. This suggests that a stochastic metamodel is more appropriate. If we now apply the additional assumption, that attrition is modelled as a Markov attrition process of the type examined by Karr [20], the probability of a change in the state (force strength) can be written in the limit as $\Delta t$ tends to 0 as:

$$P\left\{X_{t+\Delta t} = y \middle| X_t = x\right\} = A(x, y)\Delta t + o\left(\Delta t^2\right) \qquad (18)$$

The form of the infinitesimal generator matrix only allows transitions to states whose strengths differ from the initial state by 1. The assumption of a small $\Delta t$ can be interpreted as equivalent to the probability of more than one state transition in any given time interval is small, and in the first instance can be neglected.

Given the heuristic connection between the space and time intervals of section 2.1, we can choose those intervals small enough such that both assumptions constraining the values of $\Delta t$ and $a$ are met.

In any term in the sums of Equation 17 in which there is no transition, then $\Delta B$ or $\Delta R$ is 0 and Equation 17 reduces to:

$$\sum_B B = \left( \frac{B_0}{R_0} \right)^Q \sum_R R \qquad (19)$$

where the sums run between the initial and final force strengths. The sums are easily evaluated, resulting in:

$$\ln\left( \frac{\left(B^2(t) + B(t)\right) - (B_0^2 + B_0)}{\left(R^2(t) + R(t)\right) - (R_0^2 + R_0)} \right) = Q \ln\left( \frac{B_0}{R_0} \right) \qquad (20)$$

Furthermore, by relaxing the assumption that the value of $k$ is about the same for the Red and Blue forces, and parameterising their values using:

$$k_R = r k_B \qquad (21)$$

equation 20 can be re-written as

$$\ln\left( \frac{\left(B^2(t) + B(t)\right) - (B_0^2 + B_0)}{\left(R^2(t) + R(t)\right) - (R_0^2 + R_0)} \right) = Q \ln\left( \frac{B_0}{R_0} \right) + 0.5 D_R \ln r \qquad (22)$$

The similarity between this equation and the relationship developed by Helmbold [21] to describe his analysis of real-world air and land combat results, and used by Hartley [22] in his work on combat modelling, is obvious. In the cases considered by those authors, the forces strengths are sufficiently large such that the difference between $B^2$ and $B(B+1)$ would not be observed.

## 6 CASUALTY TIME SERIES AND POWER SPECTRA

We have shown already that (making the simplest assumption possible) our metamodelling approach predicts a Power Spectrum for the time series of casualties of the following form;

$$| F(B)(f) |^2 = F\{\mathrm{corr}(B(\Delta t))\}(f) \propto | f |^{-(D+1)}$$

where *F* is the Fourier transform of the time series *B* at frequency *f*. The left hand side of this equation is the power spectrum of the time series of Blue casualties *B(t)*. This should vary with frequency *f* in the form of a power law relationship. In this relationship, the exponent (*D*+1) is proportional to the fractal dimension of the locations of Red casualties on the battlefield. Since the Red fractal dimension *D* lies in the range 0 to 2, our theory predicts that this exponent should lie in the range -1 to -3.

We can rewrite the equation as;

$$P_B(f) = a|f|^{-(D+1)},$$

where *P* is the power spectrum of Blue casualties *B* at frequency *f* and *a* is the constant of proportionality. Taking logarithms, this becomes,

$$\log P_B(f) = \log a - (D+1)\log|f|.$$

In other words, when the power spectrum is plotted on a Log-Log plot, we expect a straight line with gradient -(*D*+1). As explained above, the metamodel predicts that this gradient will be between -1 and -3. To test this, a variety of MANA scenarios were run and the time series representing Blue casualty data were recorded. We then looked for evidence of a power law relationship between the power spectrum of this data *P*(*f*) and the frequency *f*.

### 6.1 MANA Senarios Used

The data required for this part of the work were time series of casualty data, i.e. the number of agents killed at each time step. MANA can record such step-by-step data, however in our work we have used "contact data", as discussed below.

Several MANA scenarios were used in order to analyse different types of situations. These scenarios were:

'Western Front in World War 2' (developed by Lauren),and the 'Lanchester', 'Meet', and 'Swarm' scenarios developed by Witty.

Contact data measures the number of detections made, over a time period. We have used contact data because there is not enough information from casualty data to calculate a good power spectrum. For example, quite often, there are no casualties for several time steps. This is because it is (for example) the companies that are modelled, not the individual men, and therefore the probability of killing an agent (a company) is small. However, we assume that every time there is contact between agents (i.e. they can detect one another), there will be casualties (perhaps only on a small scale). Therefore, contact data can be used to represent casualties. This use of contact data also replicates the approach taken in previous work. Figure 6 shows the plots of casualties (i.e. contacts) over time from these various scenarios, illustrating the large dynamic range over which we are performing this analysis.

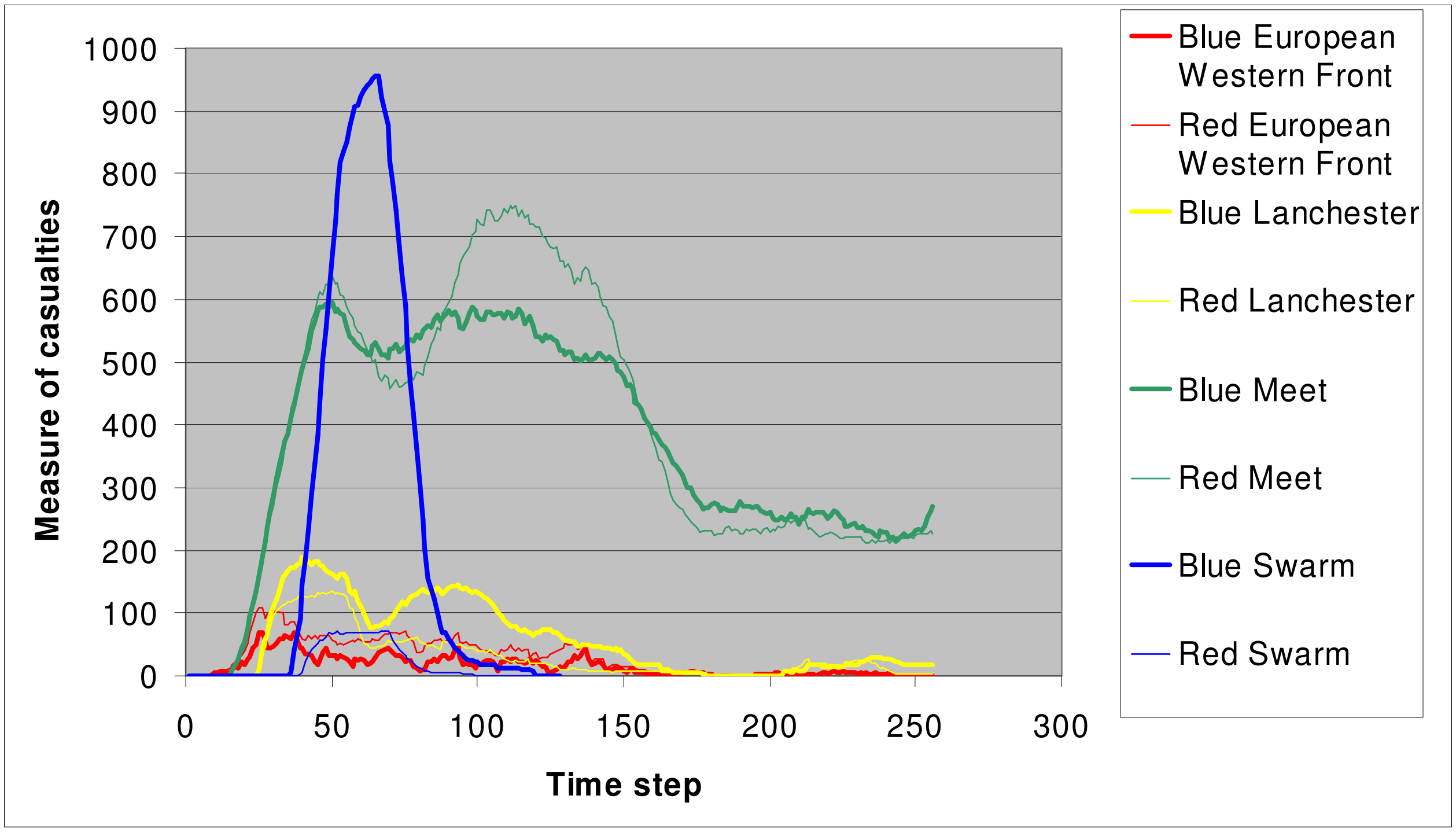

**Figure 6: Casualty time series considered in our analysis.**

### 6.2 Power Spectrum Analysis

Below is an explanation of the analysis undertaken to analyse the power spectra of the casualty time series depicted in Figure 6. A program was written specifically to carry out this analysis both for single casualty data sets, and for many time series at a time.

Firstly, a power spectrum is defined to be the squared modulus of the Fourier transform. When using discrete data, the discrete Fourier transform is used, and the squared modulus is found for each of the elements of the Fourier transform. Using sampled (discrete) data rather than the 'actual' continuous data means that we can only estimate the power spectrum. There is a limit to the number of data points that can be produced, as well as the length of the time step. Moreover, the data itself (either casualties or contacts) is discrete. When using the discrete Fourier transform, there are standard methods for overcoming the problems of 'aliasing' and 'frequency leakage' that are associated with this situation.

In our work we have sought to improve the estimate by:

- Averaging over a partitioning of the data, in which each segment has had its power spectrum taken;
- Using a window function to reduce 'leakage'. However experimental evidence suggests that this may distort the gradient of the power spectrum, so in the event this was not applied to our data;
- Replicating MANA runs and aggregating the resulting power spectra.

### 6.3 Results

**<u>Self-Organisation</u>.** Four graphs follow in Figures 7 to 10. These depict the Log-Log plots of the power spectrum of the contact data from a single MANA run in each of our scenarios. The graphs illustrate (using regression analysis) that a power law (indicated by a straight line) explains 88% to 96% of the data in our experiments. In addition, the gradients of the straight lines are between -1 and -3, as expected from our theory. These provide conclusive evidence that the casualty time series generated by the MANA model have an emergent power law distribution for the power spectrum of casualties. This is a signal of self-organisation in the MANA model.

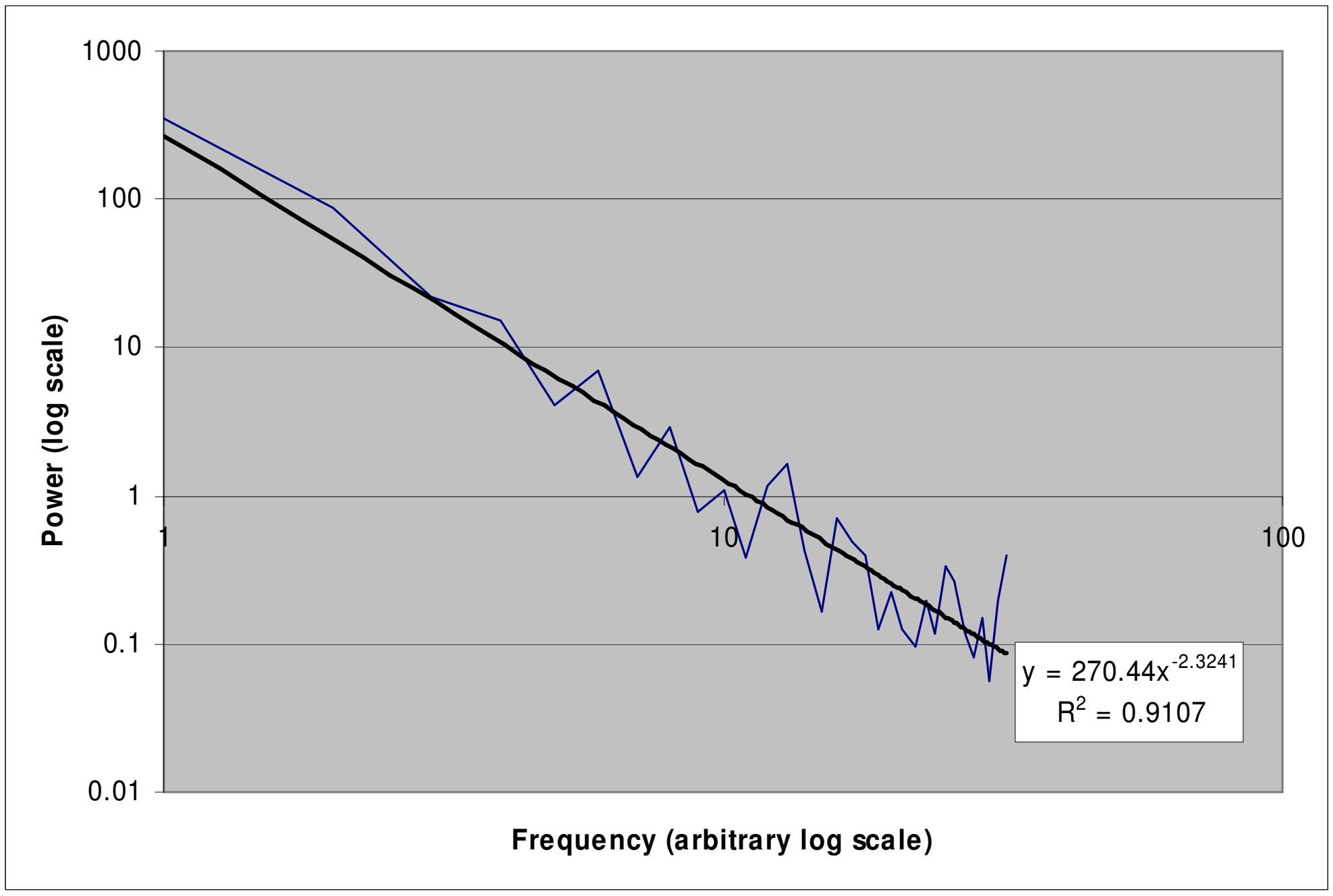

**Figure 7: Power spectrum of contact data from Lauren's European Western Front Scenario with line of best fit.**

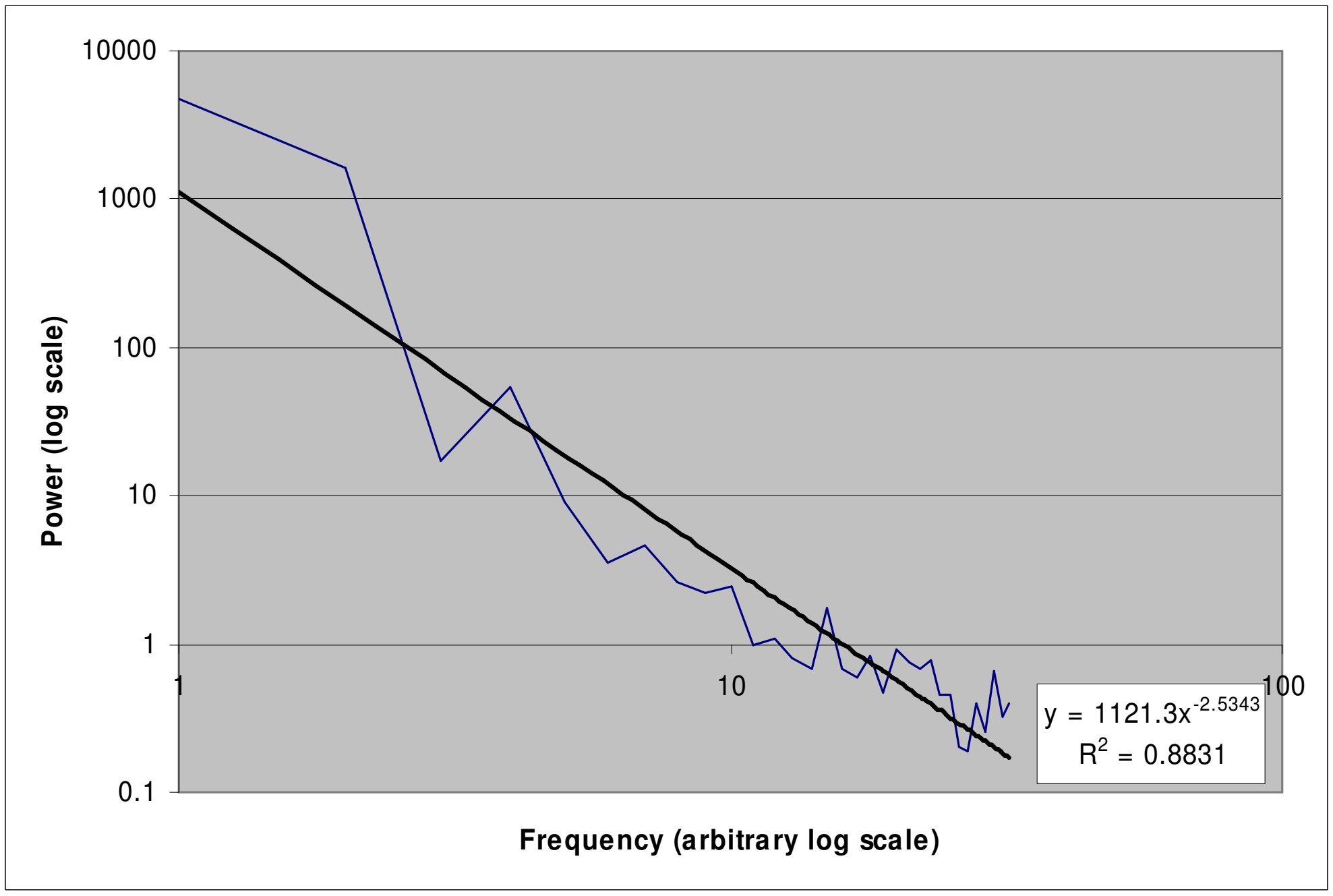

**Figure 8: Power Spectrum of casualty data from 'Lanchester' scenario with line of best fit.**

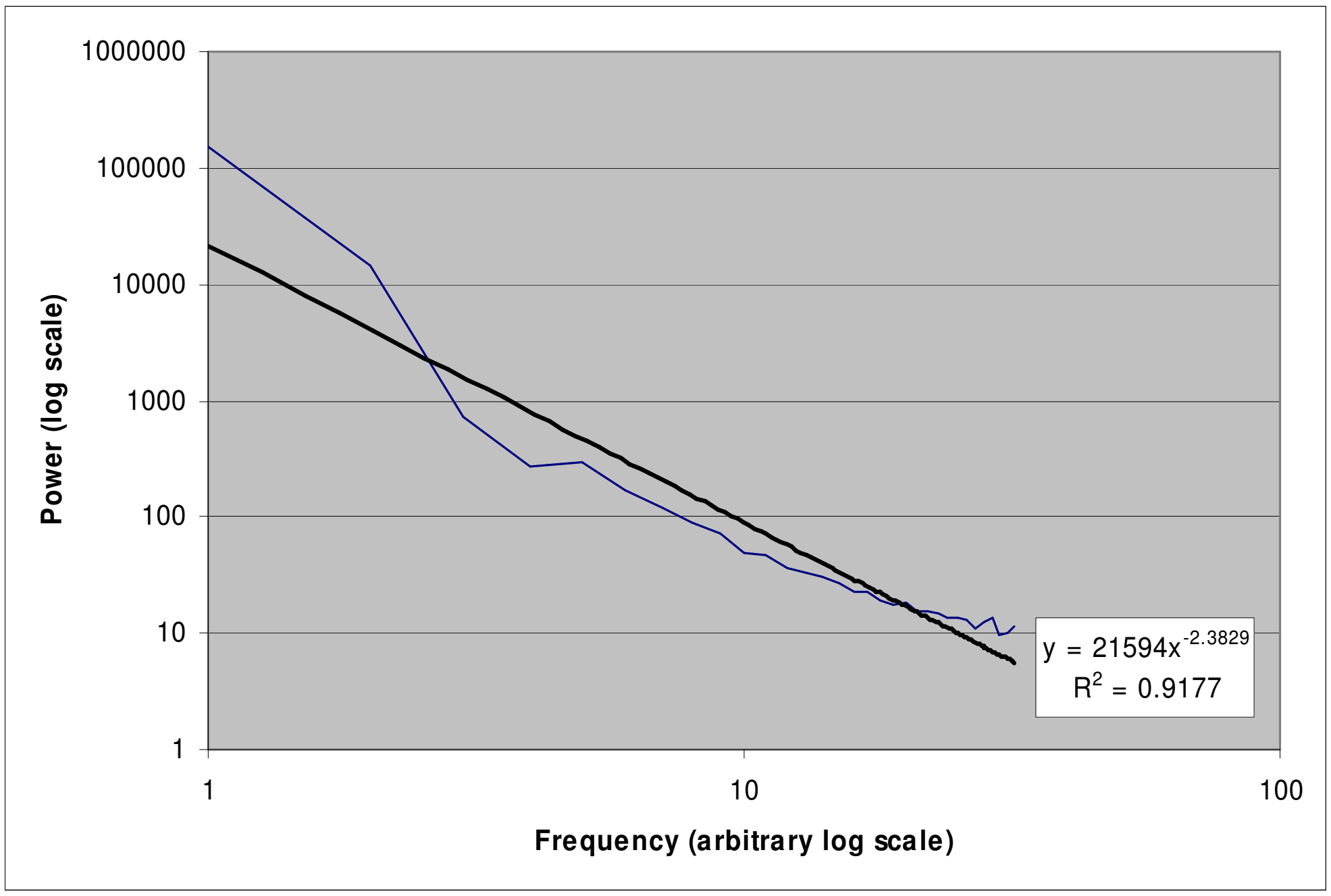

**Figure 9: Power spectrum of contact data from 'Meet' scenario with line of best fit.**

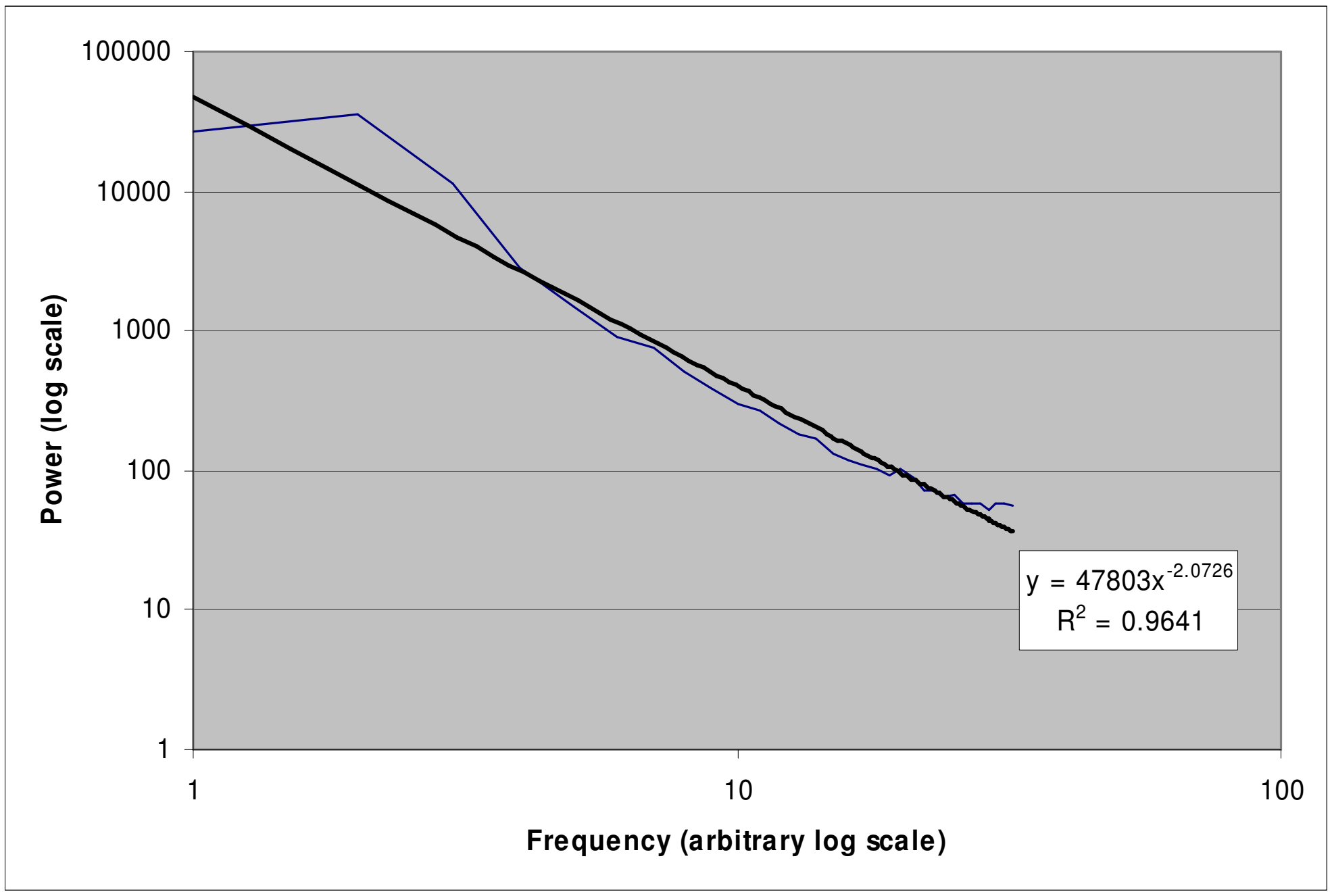

**Figure 10: Power spectrum of contact data from 'Swarm' scenario with line of best fit.**

In order to reduce the variance of the estimate further, we have replicated the MANA scenarios hundreds of times so that the power spectra of these runs can be averaged, to produce a better estimate. The results of this analysis are displayed in Figures 11 and 12.

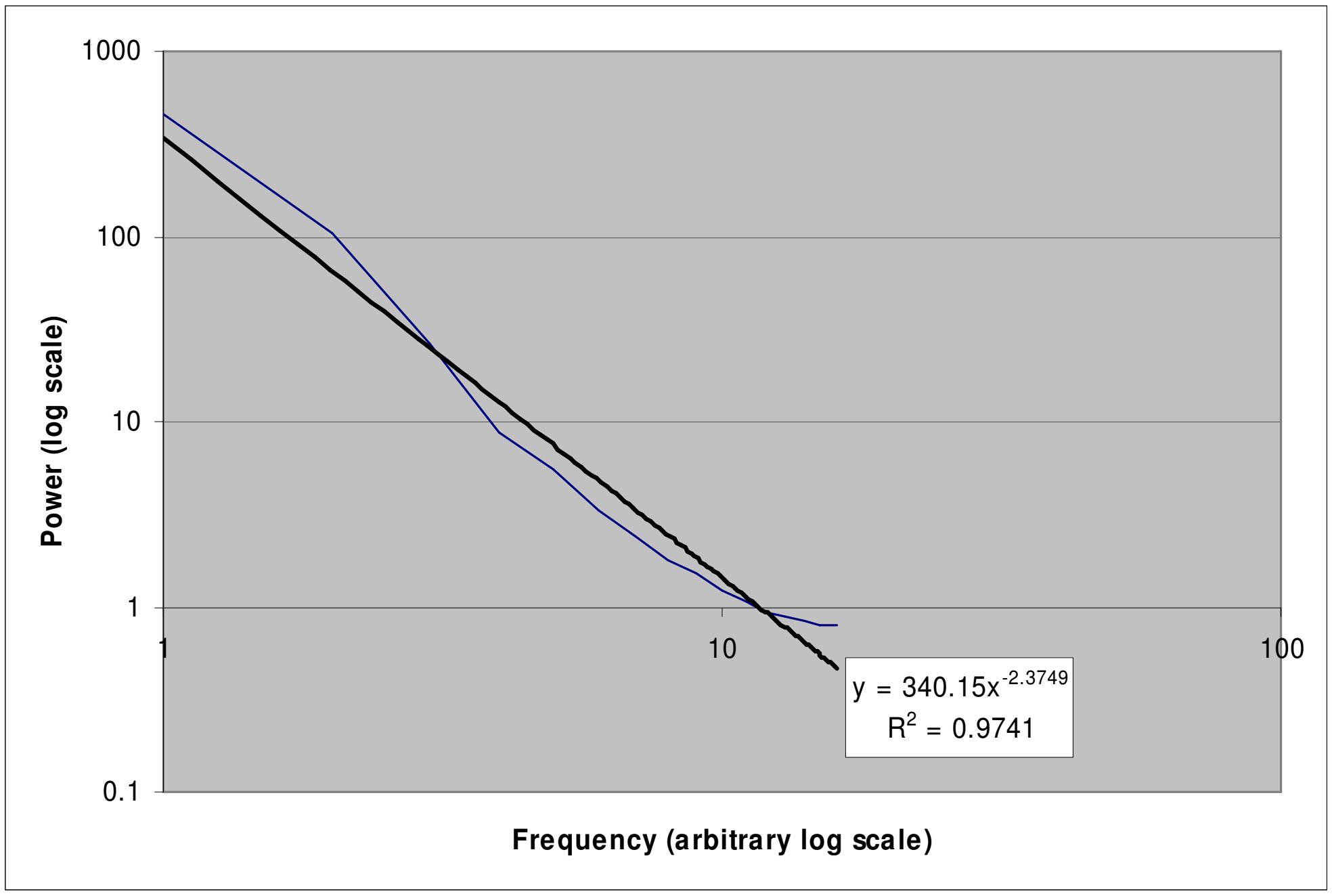

**Figure 11: Aggregated results from 255 replications of European Western Front scenario with line of best fit.**

It can be seen that a straight line explains 97% of the data in the aggregated case in Figure 11. This is a better fit than the individual replication shown in Figure 7 in which a straight line explained 91% of the data. This result confirms the gradient of about -2.3 that was seen in the single replication.

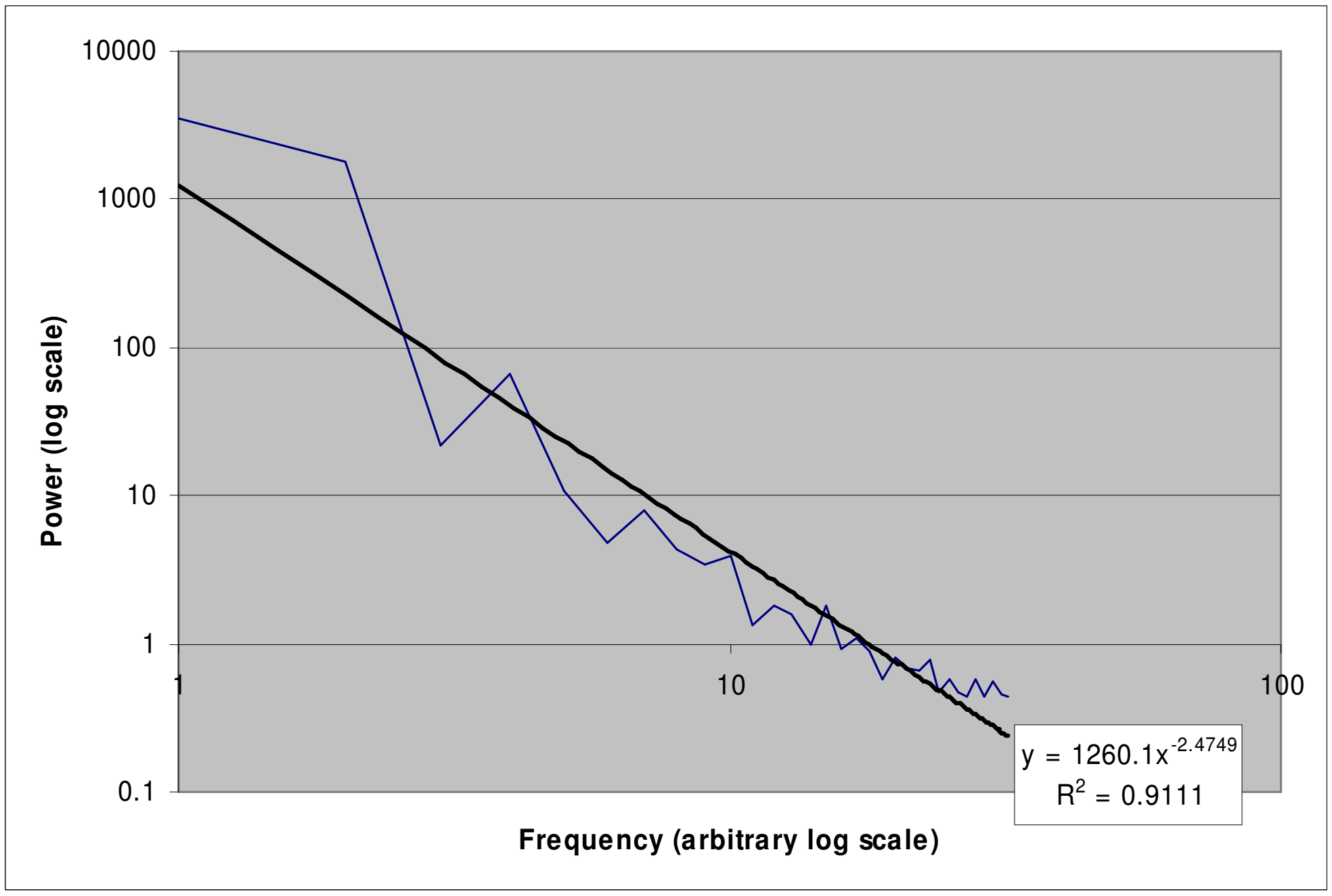

**Figure 12: Aggregated results from 250 replications of 'Lanchester' scenario power spectra with line of best fit.**

In Figure 12, 91% of the data is now explained by a straight line, compared to 88% in the single replication - Figure 8. A gradient of about –2.5 is seen, which confirms the result from Figure 8.

Figures 11 and 12 show that over many replications of these scenarios, the power law relationship is clear. and the gradient is still between –1 and –3, as expected.

### 6.4 Implications – The Volatility of Casualty Time Series

The fact that the time series of casualties has a power spectrum characterised by a power law distribution, means that the power spectrum is of the form

$$\textit{Power (squared amplitude of contribution of frequency f)} \propto f^{-\beta}$$

Thus the entire power spectrum can be characterised by a single number – the 'exponent' $\beta$. This is a characteristic of intermittent time series such as those of fractional Brownian Motion form. If we assume that the time series of casualties is 'stable' (i.e. has a roughly constant mean value over a period of time) and is of fractional Brownian Motion form, then this exponent will naturally lie in the range 1 to 3 (which again squares with our original metamodel, and with the MANA results we have shown here). The exponent can then be used to characterise the *volatility* of the time series of casualties. In fact, we have the relationship;

$$\beta = 2H + 1$$

where $H$ is the *Hurst Exponent* of the casualty time series (see Reference [11] for a discussion of the Hurst Exponent). The Hurst Exponent always takes values between 0 and 1, and we can see from this that the exponent $\beta$ will always lie between 1 and 3. If we think of the time series of casualties as an interface between the space above the time series, and the space below the time series, then this interface forms a fractal, with an interface fractal dimension *D(I)*. From this we can derive the relationships;

$$H = 2 - D(I)$$

$$\text{and hence } \beta = 5 - 2D(I)$$

*D(I)* always takes values between 1 and 2, so we see again that $\beta$ should lie between 1 and 3.

When the time series of casualties is smoother, the interface fractal dimension *D(I)* and the Hurst Exponent *H* are both closer to one. Thus the exponent of the power spectrum ($\beta$) is closer to 3. When the time series is more intermittent, *D(I)* is closer to 2, the Hurst Exponent is closer to 0, and $\beta$ is closer to 1.

Assuming that the conditions of conflict do not change, it is possible, for casualty time series having a power law Power Spectrum such as we have here, to make a short term forecast for the time evolution of casualties without the need to assume that errors are normally distributed [23].

## 7 CONCLUSIONS

Experiments with simulation runs suggest that Equation 10 provides a useful initial metamodel for the behaviour of the MANA cellular automaton combat model. This suggests that the Fractal Attrition Equation may be useful for describing scenarios where at least one side has a complicated, dispersed spatial distribution. It also provides a framework for understanding combat as a self-organising system. It thus provides a theoretical underpinning to describe why dispersed forces, guerrilla and swarming tactics can be effective in certain circumstances, as well as provide insight into the potential effects of new military paradigms such as Network Centric Warfare. For example, is the true value of networking dispersed units together that it allows the networked force to more ably maintain a favourable fractal dimension relative to its opponent?

What was particularly interesting for the Meet scenario discussed in detail here was that the experimental MANA runs revealed that the LER maintained a value of around 1 for several different force ratios. It was also seen that the fractal dimension of each side changed as this force ratio changed, as would be expected according to the Metamodel if the LER was to remain constant.

While the Metamodel was capable of describing this phenomenon, the experimental results were at odds with the classical Lanchester theory. According to this, the LER ratio should change as the force ratio changes, because the potential firepower of one side has changed relative to the other. The Lanchester theory fails to describe the observed behaviour because it does not incorporate spatial distributions.

Importantly, Equation 10, together with the additional assumptions that casualties can be modelled as a Markov attrition process and that the forces spatial distributions are fractal in nature, allows a relationship between the initial and final forces strengths to be determined that has been shown to be consistent with historical results.

One of the key predictions of our approach is that the power spectrum of the time series of casualties should be a power law. This is conclusively demonstrated from analysis of MANA data and has direct consequences for the nature of such casualties, It implies that they will tend to arrive in bursts, as an intermittent time series, rather than smoothly varying.